\documentclass[useAMS,usenatbib]{mn2e}

\usepackage[dvips]{graphicx}
\usepackage{wrapfig}
\usepackage{amsmath}

\title[The impacts of UV feedback on galaxies during the EoR]
{The impacts of ultraviolet radiation feedback on galaxies during the epoch of reionization}

\author[K. Hasegawa and B. Semelin]
{{Kenji Hasegawa $^{1,2}$ \thanks{E-mail: hasegawa@ccs.tsukuba.ac.jp}}
and {Benoit Semelin $^{2,3}$}
\\
$^{1}$Centre for Computational Sciences, University of Tsukuba, Ten-nodai, 1-1-1 Tsukuba, Ibaraki 305-8577, Japan\\
$^{2}$LERMA, Observatoire de Paris, CNRS, 61 Av. de l'Observatoire, 75014 Paris, France \\
$^{3}$Universit$\acute{e}$ Pierre et Marie Curie, 4 place Jussieu, 75005 Paris, France}
\begin{document}

\date{Accepted  Received ; in original form }

\pagerange{\pageref{firstpage}--\pageref{lastpage}} \pubyear{}

\maketitle

\label{firstpage}

\begin{abstract}
We explore the impacts of ultraviolet (UV) radiation feedback on galaxies during the epoch of reionisation by cosmological simulations in which hydrodynamics and the transfer of the H and He ionising photons are consistently coupled. Moreover we take into account $\rm H_2$ non-equilibrium chemistry, including photo-dissociation. The most striking feature of the simulations is a high spatial resolution for the radiative transfer (RT) calculation which enables us to start considering not only external UV feedback processes but also internal UV feedback processes in each galaxy. We find that the star formation is significantly suppressed due to the internal UV and supernova (SN) feedback. In low mass galaxies with $M<10^9M_{\odot}$, a large amount of gas is evacuated by photo-evaporation as previous studies have shown, which results in the suppression of star formation.  Surprisingly, star formation in massive halos is also strongly suppressed despite the fact that these halos hardly lose any gas by photo-evaporation. The suppression of star formation in massive halos is mainly caused by following two factors; (i) small scale clumpy structures in the galaxies are smoothened by the internal feedback, (ii) although the dense gas in the galaxies is mostly neutral, the $\rm H_2$ formation and cooling processes are disturbed by mild photo-heating. Photo-dissociating radiations actually suppress star formation, but the magnitude of the effect is not so large in massive galaxies. Even though our simulation volume is too small to be a representative patch of the Universe during reionisation, we find that our simulated star formation rate densities and HI fractions at $z\sim6-7$ are consistent with those found in observations. 
%However none of our simulated reionisation histories are able to match the optical depth for the Thomson scattering of CMB photons measured by WMAP. These results might indicate that contributions of missing sources, such as Population III stars in mini-halos and/or back holes, are required to reionise the early Universe.  

\end{abstract}

\section{Introduction}
Many observations have indicated that the Universe was reionised early on. 
According to the Gunn-Peterson effect \citep{GP65} in the spectra of high-$z$ quasars, the Universe was already reionised by $z\sim6$ \citep{Fan06a,Fan06b}.  
In addition, the {\it Wilkinson Microwave Anisotropic Probe} (WMAP) satellite produced further information on the reionisation epoch. 
The measured optical depth for the Thomson scattering of CMB photons by free electrons is $\tau_{\rm e}=0.087\pm0.014$. 
Assuming an instantaneous reionisation as a toy model, this value indicates that the Universe was reionised at $z=10.4\pm1.2$  \citep{Jarosik11}. 
%Moreover, in the last decade, the evolution of the luminosity function of Lyman-$\alpha$ emitters (LAEs) has also been used to evaluate the neutral hydrogen fraction during the epoch of reionisation \citep{Ouchi09}. 
Moreover, in the last decade, some observed quantities, such as the spectra of Gamma-ray burst 050904 \citep{Totani06} and the evolution of the luminosity function of Lyman-$\alpha$ emitters (LAEs) \citep{Ouchi09}, have also been used to evaluate the neutral hydrogen fraction during the epoch of reionisation.
However the exact reionisation history is still unclear, since we cannot directly detect the ionisation state of intergalactic medium (IGM) even with the latest observational facilities. 

%One of the way to probe the undetected properties of IGM is the approach by numerical simulations. 
Numerical simulations are means of increasing our understanding of the physical properties of the undetected IGM. 
Many numerical simulations of the cosmic reionisation have been performed \citep{Nakamoto01,Ciardi03, Sokasian04,Iliev06b,Iliev07,TC07,Finlator09,AT10,PS11,Ahn12}. 
Moreover, in order to predict the 21-cm signal emitted in intergalactic medium (IGM) during the reionisation epoch, which should be observed by radio-interferometers such as LOFAR, MWA and SKA, simulations of the cosmic reionisation have been performed. \citep{CM03,Mellema06,Lidz08,Santos08,Thomas09,Baek09,Vonlanthen11}. 

The nature of the sources of the cosmic reionisation is still uncertain, but stars in galaxies, at least, are expected to contribute to the process, since many galaxies have already been observed at $z>6$.  
In a cold dark matter (CDM) cosmology, structure formation proceeds in a bottom-up manner. 
Therefore it is expected that low mass galaxies were dominant sources of ionising photons during the early phase of reionisation. 
However it is also well known that star formation in low mass galaxies is very sensitive to feedback effects owing to their shallow gravitational potential. 
For instance an external ultraviolet background (UVB) radiation field which develops as the cosmic reionisation proceeds heats the gas up to $\sim 10^4$K \citep{TW96, Okamoto08} and dissociates molecules \citep{Ricotti01, SU04, HUK09,Johnson12}. 
Moreover \cite{Susa08} have shown an external UVB can suppress the star formation even in a galaxy that is massive enough to 
sustain an ionised hot gas. The suppression in the massive galaxy is mainly due to the fact that photo-heating prevents gravitational inability. 
Not only the external UVB but also UV feedback from its own stars is likely to regulate star formation in each galaxy. 
In order to investigate ionising photon escape fraction from high$-z$ dwarf galaxies, \cite{WC09} have performed radiation hydrodynamics (RHD) simulations in which the internal UV and SN feedback are included. 
They have shown that the internal feedback evacuates a large amount of gas from  halos with masses less than $10^{8.5}M_{\odot}$, and regulates the star formation rate. 
If previous episodes of star formation and feedback effects associated with the star formation, which are not included in their simulations, are taken into account,  the impact of the internal feedback is expected to be more drastic. 
In addition, such evacuation would lead to the enhancement of the escape fraction of ionising photons, which is an important factor to regulate reionisation history. 
According to the previous studies, it is strongly expected that UV feedback plays a very important role in regulating the histories of the cosmic star formation and the reionisation. 
But unfortunately it is often hard to implement the internal UV feedback in simulations of the cosmic reionisation, since calculation costs for the RT are very high.

\cite{Iliev07} have explored the reionization history regulated by UVB with high resolution $N$-body and post-processing RT calculations,  
assuming a  constant characteristic halo mass of $10^9M_{\odot}$ below which stars are forbidden to form in ionised regions. 
However, this characteristic mass should depend on the halo formation epoch, the strength of the UVB radiation field, and so on \citep{Kitayama01, SU04,Okamoto08}. 
Moreover the internal UV feedback cannot be correctly included due to a lack of spatial resolution in RT calculations, although the spatial and mass resolutions for $N$-body simulations are sufficiently high to take into account ionising photons from small structures. 
Hence the star formation history in their simulations might be inaccurate. 
\cite{Finlator11} have recently performed simulations of the cosmic reionisation in which the transfer of UV photons is consistently coupled to hydrodynamic calculations. 
Hence they can consider suppression of star formation in galaxies due to photo-evaporation without assuming a characteristic mass. 
However the spatial resolution for their RT calculations is not high enough to correctly take into account the internal UV feedback. 
The lack of spatial resolution also implies using an escape fraction parameter, which is frequently assumed to be constant over all masses of galaxies. 
Since the ionising photon escape fraction likely depends on various properties of the galaxy, it is doubtful that the assumption of a constant value is valid.  
In addition, there is little agreement on the value of the escape fraction in previous studies \citep{Gnedin08, WC09, RS10, Yajima11}.  
Hence the employment of the escape fraction results in a large uncertainty of the reionisation history. 
\cite{PS11} have implemented a moment-based RT solver in {\small GADGET}, and have simulated the cosmic reionisation. With the code, spatial resolutions for the RT and hydrodynamics calculations are equivalent so that the internal UV feedback can be appropriately considered. In addition, they did't have to employ a "galactic" escape fraction. 
However there might be still an unresolved interstellar medium (ISM) where some photon absorption is expected. 
Hence they adopted an "ISM" escape fraction to take into account the photon losses in the unresolved ISM. 
Although some of their models well produce a plausible history of reionisation, the minimum halo mass resolved in their simulations,  typically $3.0\times10^9 h^{-1} M_{\odot}$ but $3.8\times10^8 h^{-1} M_{\odot}$ in the maximum resolution run, seems too massive to discuss the suppression of star formation at the epoch of reionisation \citep{Okamoto08}. 
Therefore the impacts of the internal and external UV feedback on the histories of star formation and reionisation have not been well established yet. 

In this paper, we aim to clarify the importances of UV feedback on cosmic reionisation history by radiation hydrodynamics simulations. 
Unlike previous numerical simulations of the cosmic reionisation, we include non-equilibrium chemistry regarding the $\rm H_2$ molecule, which is the most efficient coolant below $10^4$K in a low-metallicity environment. 
$\rm H_2$ cooling plays a crucial role not only for less massive halos with $T_{\rm vir}<10^4$K but also for massive halos with $T_{\rm vir}>10^4$K. 
Needless to say that halos with $T_{\rm vir}<10^4$K cannot collapse without $\rm H_2$ cooling in the early Universe. 
But even in halos with $T_{\rm vir}>10^4$K, internal structures would be affected by thermal processes. 
We have to trace the dynamical and thermal evolution of gas in the cold $T<10^4$K regime, since stars are born in cold and dense regions.    
In addition, resolving such internal structures is very important to understand how ionising photons escape from a galaxy \citep{FS11}. 
Very recently, \cite{Johnson12} have shown that the global and local photo-dissociating radiation regulates cosmic star formation history in the epoch of reionization. However the roles of an inhomogeneous reionisation process and the internal photo-ionisation feedback are still unclear, because they simply assumed a homogeneous reionisation model. 
We show in this paper not only photo-dissociating radiation but also photo-ionisation radiation plays crucial roles on the reionisation and star formation history.  

This paper is organized as follow. In Section 2, we describe our numerical method and the setup of the simulations. 
We show the results of the simulations in Section 3. In Section 4, we compare the simulation results with observations, and also discuss the expected effect of some processes that we did not consider in this work. Finally we devote Section 6 to the conclusions.  
Throughout this paper we assume a $\Lambda$ cold dark matter universe with cosmological parameters: $\Omega_b = 0.049$, $\Omega_0=0.27$, $\Omega_\Lambda=0.73$, $h=0.71$, and $\sigma_8=0.81$ based on 7-year WMAP results \citep{Jarosik11}

\section{Simulation}
\begin{table*}
\caption{List of runs}
\begin{center}
\begin{tabular}{ccccccc}
  \hline
  Name  & box size [Mpc] & Numbers of particles & RT & photodissociation & 
  $f_{\rm e,ISM}$ \\
  \hline
   rN256L5 & 5 & $256^3 \times 2$ & no &  - & - \\
   rN256L5RT & 5 & $256^3 \times 2$ & yes & shielding function & 1.0  \\
   rN128L5 & 5 & $128^3 \times 2$ & no & - & - \\
   rN128L5RT & 5 & $128^3 \times 2$ & yes & shielding function & 1.0 \\
   rN128L5RT-fe08 & 5 & $128^3 \times 2$ & yes & shielding function & 0.8 \\
   rN128L5RT-fe06 & 5 & $128^3 \times 2$ & yes & shielding function & 0.6 \\
   rN128L5RT-fe04 & 5 & $128^3 \times 2$ & yes & shielding function & 0.4 \\
   rN128L5RT-fe02 & 5 & $128^3 \times 2$ & yes & shielding function & 0.2 \\
   rN128L5RT-noLW & 5 & $128^3 \times 2$ & yes & no & 1.0 \\  
   rN128L5RT-nfsh & 5 & $128^3 \times 2$ & yes & optically thin & 1.0 \\
   \hline
\end{tabular}
\end{center}
\end{table*}
\begin{figure}
	\centering
	{\includegraphics[width=8cm]{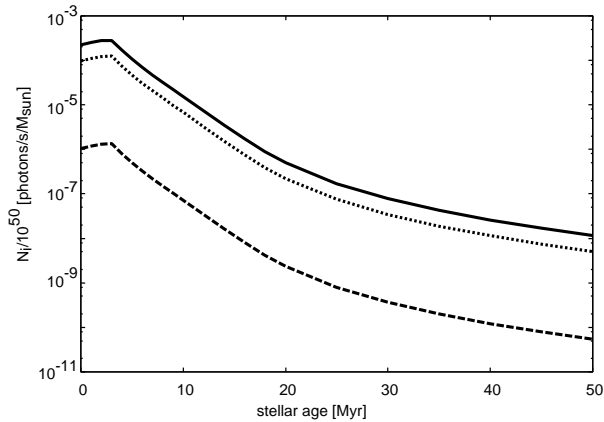}}	
	\caption{Specific ionising photon production rate as a function of stellar age. The solid, dotted, and dashed lines 
	respectively show the photon production rates of HI, HeI, and HeII ionising photons.}
	\label{rate}
\end{figure}
In order to self-consistently simulate the reionisation and star formation history, we have to couple hydrodynamics to the transfer of UV photons from numerous radiation sources. 
For this purpose we use a novel radiation hydrodynamics code {\small START} (SPH with Tree-based Accelerated Radiative Transfer) \citep{HU10}, which enables us to solve such a numerically expensive problem. 
We describe the detailed simulation method in the following part. 

\subsection{Feedback Processes}
\subsubsection{Star Formation Recipe}\label{SFrecipe}
We adopt the star formation recipe employed in \cite{SU04} who explored the impact of UVB on the formation of dwarf galaxies. 
We allow each gas particle to form stars, if a gas particle satisfies the following three conditions; (i) $\rho > 200\rho_{\rm av}$, (ii) $T < T_{\rm cr}=5,000$K, and (iii) ${\rm div}{(v)} <0$. 
The stars are born at a rate of 
\begin{equation}
	\frac{d \rho_*}{dt} = C_*\frac{\rho_{\rm gas}}{t_{\rm dyn}}, 
\end{equation}
where $\rho_*$, $\rho_{\rm gas}$, and $C_*$ are the star and gas mass density, and the star formation efficiency, respectively. The dynamical time $t_{\rm dyn}$ is evaluated as $t_{\rm dyn}=1/\sqrt{4\pi G \rho}$. Here the condition (ii) is the most crucial, since this condition 
can be satisfied only if the gas is well shielded against UV radiation so that $\rm H_2$ cooling effectively works. However the change of $T_{\rm cr}$ hardly affects the final results as long as $T_{\rm cr}$ is well smaller than 10,000K, since the cooling time is well shorter than the dynamical time in a self-shielded region \citep{HUK09}. 
In previous simulations of galaxy formation, $C_*\sim0.01$ is frequently employed so as to reproduce the Schmidt-Kennicutt relation\citep{NS00,Saitoh08}. In addition, \cite{SU04} have shown the final stellar mass fraction in a halo is almost independent on the value of $C_{*}$ in the case where the star formation is regulated by UV feedback, although the star formation history is slightly delayed for smaller value of $C_{*}$. Therefore we carry out all of the simulations with $C_*=0.03$.

\subsubsection{SN feedback}
We include thermal feedback by SN explosions. 
Firstly, in advance of the simulations, we compute population synthesis using {{\small ${\rm P\acute{E}GASE}$}} \citep{FV97}. 
Assuming a Salpeter IMF with masses of $1M_\odot$-$100M_\odot$ and an instantaneous starburst, we obtain a specific SN rate  as a function of a stellar particle age. 
According to the rate, we feed the neighboring 50 particles around the star particle with SN energy in thermal form, assuming a released energy of $10^{51} {\rm erg}$ for one SN explosion. 

Chemical feedback is not included in this study. The inclusion of metal enrichment enhances cooling rate. In addition, dust grains likely play an important role; $\rm H_2$ formation on the surface of dust grains would increases $\rm H_2$ fraction and the absorption of UV photons by the grains leads to  a decreased UV photons escape fraction. We will include chemical feedback in a forthcoming paper. 

\subsubsection{UV Feedback}\label{feedback}
\begin{figure*}
	\centering
	{\includegraphics[width=15cm]{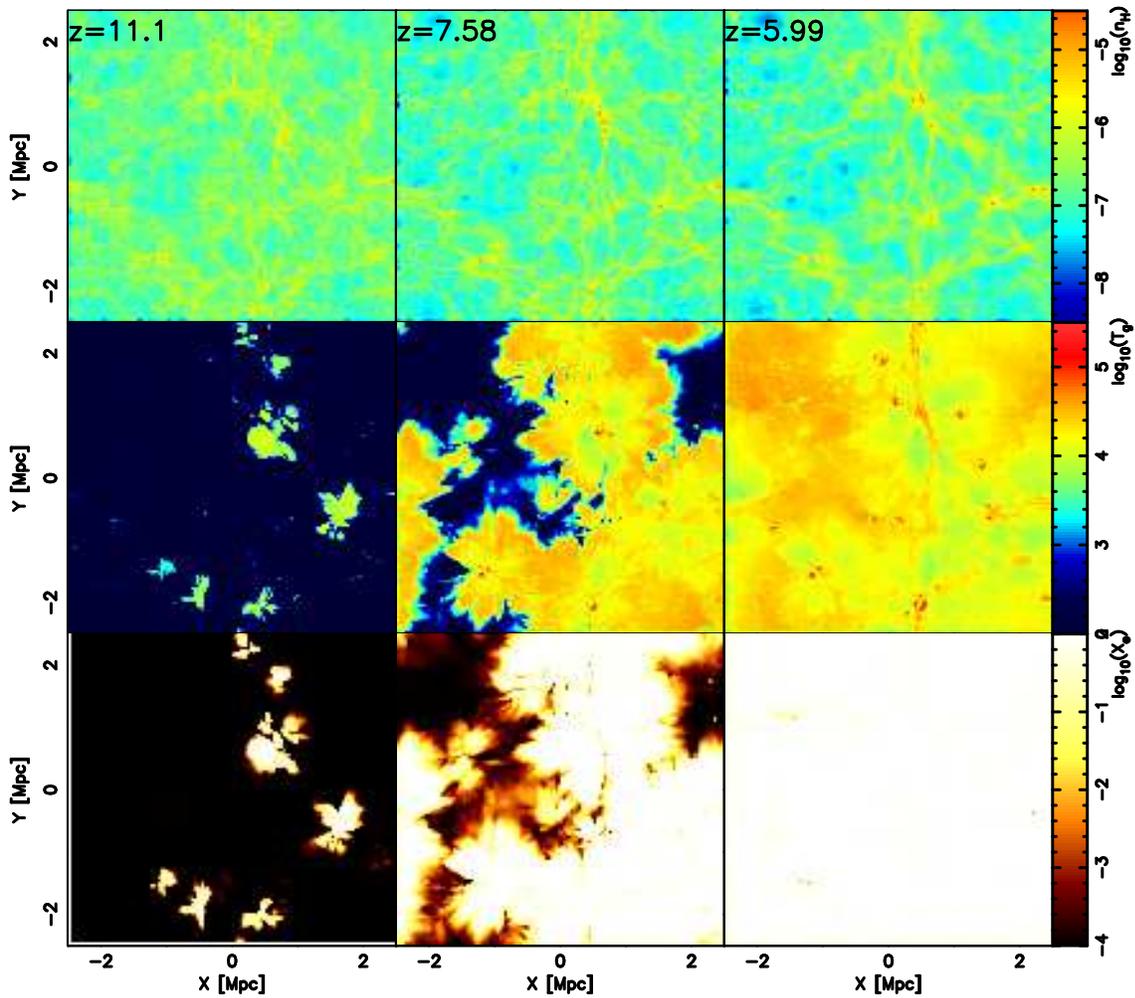}}	
	\caption{Maps of physical quantities in a slice cut through the mid-plane of the simulation box at 
	redshifts $z=11.1$, $7.58$, and $5.99$ from left to right. The evolution of comoving number density, 
	temperature, and ionisation degree is shown in the upper, middle, and lower rows, respectively.}
	\label{map}
\end{figure*}
We use the RHD code {\small START}. 
Here we briefly describe the RT method implemented in the code. 
The implementation of ray-tracing in {\small START} is basically the same as that adopted in {\small RSPH}, which has been used for simulations of structure formation in the early Universe \citep{SU06,Susa07,Susa08,HUS09,SUH09}. 
Since the rays are cast from all radiation sources to all SPH particles, the {\small RSPH} scheme allows us to solve RHD problems very accurately \citep{Iliev06a, Susa06,Iliev09}. 
Moreover, this ray-tracing algorithm enables us to solve RT without losing the spatial resolution of hydrodynamics calculation, since SPH particles are directly used for RT grids. 
With {\small RSPH}, however, the calculation cost is proportional to a number of radiation sources, and it becomes difficult to solve problems in which numerous radiation sources should be considered. 
In order to avoid this difficulty, a tree-based acceleration algorithm is employed in {\small START}. 
In the tree-based acceleration algorithm, we first make an oct-tree structure \citep{BH86} for radiation sources.  
To make the oct-tree structure, we create the root cell which contains all radiation sources, and recursively divide the cell into eight cells down to the level where each cell contains one radiation source. 
Then we check if a cell is distant enough from a target SPH particle by using following criteria;
\begin{equation}
	\frac{l_{\rm s}}{d_{\rm s}}<\theta_{\rm crit}, 
\end{equation} 
where $d_{\rm s}$ is the distance between the cell and the target SPH particle, $l_{\rm s}$ is the size of the cell, and $\theta_{\rm crit}$ is the tolerance parameter. If the above condition is satisfied, the radiation sources in the cell are regarded as one luminous virtual source. Adopting this algorithm, the calculation cost is dramatically reduced. We have to adopt a sufficiently small value of  $\theta_{\rm crit}$ to obtain accurate results.  $\theta_{\rm crit}=0.5$ is a reasonable choice (\cite{HU10}, see also \cite{Okamoto11}). 

We consider the photo-ionisation of HI, HeI, and HeII by solving the transfer of ionising photons appropriately. 
The cross-sections are respectively given by 
\begin{equation}
	\sigma_{\rm HI}(\nu) = 6.30\times 10^{-18}\left(\frac{\nu_1}{\nu}\right)^3 {\rm cm^2},
\end{equation}
\begin{equation}
	\sigma_{\rm HI}(\nu) = 7.42\times 10^{-18}\left[1.66\left(\frac{\nu_1}{\nu}\right)^{2.05}
	-0.66\left(\frac{\nu_1}{\nu}\right)^{3.05}\right] {\rm cm^2},
\end{equation}
\begin{equation}
	\sigma_{\rm HeII}(\nu) = 1.575\times 10^{-18}\left(\frac{\nu_1}{\nu}\right)^3 {\rm cm^2},
\end{equation}
where $\nu_1$, $\nu_2$, and $\nu_3$ are the Lyman limit frequencies of HI, HeI, and HeII, respectively. 
The corresponding photo-ionisation rates are expressed as 
\begin{equation}
	k_{\rm HI} = \int^{\infty}_{\nu_1}\int \frac{I_{\nu,0}{\rm exp}({-\tau_{\nu}})}{h\nu} 
	\sigma_{\rm HI}(\nu)d\Omega d\nu, 
\end{equation}
\begin{equation}
	k_{\rm HeI} = \int^{\infty}_{\nu_2}\int \frac{I_{\nu,0}{\rm exp}({-\tau_{\nu}})}{h\nu} 
	\sigma_{\rm HeI}(\nu)d\Omega d\nu, 
\end{equation}
\begin{equation}
	k_{\rm HeII} = \int^{\infty}_{\nu_3}\int \frac{I_{\nu,0}{\rm exp}({-\tau_{\nu}})}{h\nu} 
	\sigma_{\rm HeII}(\nu)d\Omega d\nu. 
\end{equation}
As shown in previous studies \citep{TBN86, FM94, Nakamoto01}, integrated photoionisation rates can be obtained knowing the optical depths at the three Lyman limit frequencies only, since we already know the shape of intrinsic intensity $I_{\nu,0}$ and the dependence of the cross-sections on frequency. 
We produce a three dimensional, i.e., a function of $\tau_{\nu_1}$, $\tau_{\nu_2}$, and $\tau_{\nu_3}$, look-up table by integrating photo-ionisation rates over all frequency ranges as a preprocessing of the simulations. 
By using the look-up table we can appropriately take into account the dependence of their cross-sections on frequency.  
A similar method is applied to calculate photo-heating rates. 

Hydrogen molecules are easily dissociated by Lyman-Werner (LW) band (11.2-13.6eV) photons via the two step Solomon process. 
We take the photo-dissociation of $\rm H_2$ into account by calculating the column density of $\rm H_2$ and adopting the self-shielding function introduced by \cite{DB96}. Then the flux in the LW band can be written as 
\begin{equation}
F_{\rm LW} = F_{\rm LW,0} f_{\rm s}
\left( N_{\rm H_2,14 } \right) \label{LW}
\end{equation}
where $ F_{\rm LW,0}$ is the incident flux, 
$ N_{\rm H_2,14}$ is the H$_2$ column density in units of 
$10^{14}{\rm cm^{-2}}$, and
\begin{equation}
f_{\rm s}(x) = \left\{
\begin{array}{cc}
1,~~~~~~~~~~~~~~x \le 1 &\\
x^{-3/4}.~~~~~~~~~x > 1 &
\end{array}
\right. \label{shield-function}
\end{equation}
We also include photo-detachment of $\rm H^-$ and photo-dissociation of $\rm H_2^+$, assuming these species are optically thin against the radiation owing to their considerably smaller abundances. 
These rates can be expressed as  
\begin{equation}
	k_{\rm i} = \int_{\nu_{\rm i}}^{\nu_{\rm 1}} \int \frac{I_{\nu}\sigma_{\rm i}(\nu)}{h\nu}d\Omega d\nu, 
\end{equation}
where $h\nu_{\rm H^-}=0.74$eV and $h\nu_{\rm H_2^+}=0.062$eV. The cross-sections are taken from \cite{Tegmark97} and \cite{Stancil94}, respectively. 
Here we should note that the self-shielding function for $\rm H_2$ photo-dissociation was derived with the assumption that a medium is static. 
Consequently the shielding effect might be stronger than in the case where velocity gradients are properly included. 
Therefore we additionally simulate two extreme cases, by changing the treatment of $\rm H_2$ photo-dissociation; (i) a run in which we never use the shielding function ( ${\rm H_2}$ molecules are optically thin in the LW band). (ii) a run in which photo-dissociation processes (not only $\rm H_{\rm 2}$ photo-dissociation but also $\rm H^-$ photo-detachment and $\rm H_2^{+}$ photo-dissociation processes) are excluded. We verify the importances of the photo-dissociation processes by performing these two runs. 

We compute the HI, HeI and HeII ionising photon production rates in {{\small ${\rm P\acute{E}GASE}$}} \citep{FV97}, in the same way that we evaluate the specific SN rate as a function of stellar age. 
We plot the photon production rates as a function of stellar age in Fig.\ref {rate}. 
Here, for simplicity, we assume that the shape of spectral energy distribution (SED) always corresponds to that of a black body spectrum with an effective temperature of 50,000K. The assumption is reasonable due to the following reasons; (i) UV radiation is dominantly emitted by young stars. (ii) With the IMF we choose, the SED of young stellar population can be well approximated by a black body with an effective temperature of 50,000K \citep{Baek09}. Evaluating reaction rates between UV radiation and matters by the way
 described above, we implicitly solve a non-equilibrium chemistry regarding nine species, i.e., $\rm e^-$, $\rm H^+$, $\rm H$, $\rm H^-$, $\rm H_2$, $\rm H_2^+$, $\rm He$, $\rm He^+$, and $\rm He^{2+}$. The initial fractions are taken from \cite{GP98}. 

Finally, there is a synergy effect between SN and UV feedback. 
It is often true that thermal feedback by SN explosions cannot work efficiently, because radiative cooling processes efficiently work in a high density regions, which are the sites of star formation, and gas in these regions turns out to be cooled quickly. 
In our simulations, a timestep is basically much smaller than the lifetime of massive stars. 
Hence UV feedback works prior to the first SN explosion. The UV radiation from  newly born stars heats the adjacent medium, and the gas density decreases before thermal energy released by SN explosions is poured into the gas. Consequently, SN thermal feedback works more efficiently when UV feedback is included. 

\subsection{Setup}\label{setup}
Each simulation starts from $z=150$ with initial conditions generated with {\small Grafic-2} \citep{Bertschinger01}. 
We performed simulations with two different resolutions. 
In high resolution runs, we handled  $256^3$ particles each for SPH and DM particles in a volume of comoving $L_{\rm box}=5 {\rm Mpc}$ on a side. 
In this situation, the particle masses correspond to $4.9\times10^4M_{\odot}$ for SPH and $2.5\times 10^5M_{\odot}$ for DM, respectively. 
Hence a small galaxy of $M_{\rm halo,min}\approx 2.5\times 10^7M_{\odot}$ can be resolved with 100 particles. 
We note that this mass resolution is $\sim20$ times better than that in the highest resolution run of \cite{PS11}. 
The gravitational softening length is set to $0.03\times L_{\rm box}/N^{1/3}$, which corresponds to 595 comoving pc. 
In low resolution runs, on the other hand, we handled  $128^3$ particles each for SPH and DM particles in the same size box. 
Hence the minimum halo mass in the low resolution run is $M_{\rm halo,min}\approx 2.0\times 10^8M_{\odot}$ which is simply 8 times higher than that in the high resolution run.  
Since the spatial resolutions of both the RT and hydrodynamics calculations are sufficiently smaller than the typical size of small galaxies, we can appropriately take into account internal UV feedback in each galaxy. 
The spatial resolution of the high resolution RT simulation is roughly equivalent to those of previous numerical works that have evaluated ionising photon escape fractions of ionising photons from galactic halos \citep{Gnedin08,RS10}. 
Although photon losses in galaxies can automatically be taken into account owing to the high spatial resolution, there still might be absorptions in the spatially unresolved ISM. 
\cite{PS11} have adopted an "ISM" escape fraction $f_{\rm e,ISM}$ to account for such unresolved absorptions. 
We note that this "ISM" escape fraction has a completely different physical meaning from the "galactic" escape fraction which has been usually utilized in previous simulations on reionisation \citep{Iliev07, Finlator11}. 
In our simulations we basically work with $f_{\rm e,ISM}=1.0$, but we also carry out simulations with lower values of the ISM escape fraction. 
We should emphasize that the variance of galactic escape fractions connected to the different properties of each halo is automatically taken into account even if we employ an ISM escape fraction. 
The validity and dependence of our results on the ISM escape fraction will be discussed in Section \ref{escape}. 
Parameter sets for all runs are summarized in Table1. 

\section{Results}
\subsection{Global Evolution}
\begin{figure}
	\centering
	{\includegraphics[width=7cm]{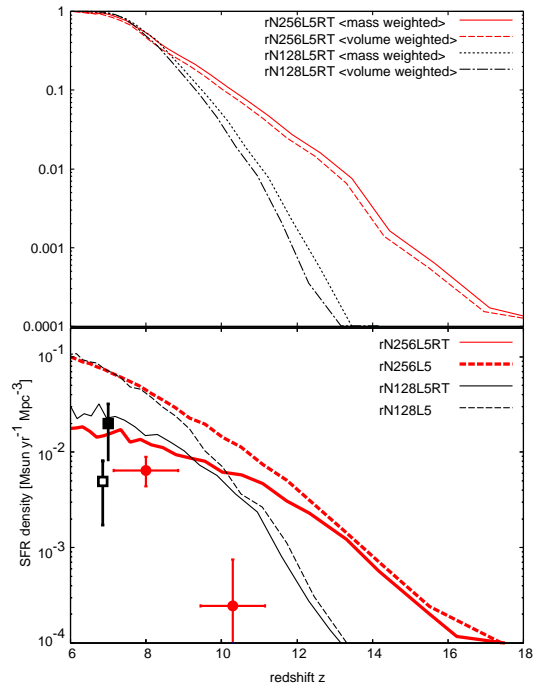}}	
	\caption{$Upper\;panel$: Hydrogen ionisation histories. 
	The solid and dashed curves respectively indicate the mass and volume weighted ionisation histories in the rN256L5RT run.
	while the dotted and dat-dashed curves respectively indicate the mass and volume weighted ionisation histories in the rN128L5RT run.
	$Lower\;panel$: Cosmic star formation rate densities. The cosmic star formation rate densities in the rN256L5RT,  
	rN256L5, rN128L5RT and rN128L5 runs are respectively shown by the thick solid, thick dashed, thin solid, and 
	thin dashed curves. 
	The filled and open squares are respectively the extinction corrected SFRD integrated 
	down to $L=0$, and the no extinction corrected SFRD integrated down to $L\approx0.1L^*$
	from Ouchi et al. (2009). 
	The filled circles are the SFRD integrated down to $0.05L^*_{z=3}$ from Bouwens et al. 2011.}
	\label{ionsfdhis}
\end{figure}
In this section, we show the evolution of global quantities, such as ionised fraction and star formation rate density. 
We would like to point out that our simulation box is too small to mitigate the effect of cosmic variance on these quantities. 
Although the presented global quantities cannot be considered to be representative of the whole Universe, these quantities are very useful to understand the roles of UV radiation feedback. In addition, these quantities can be used for the comparison with the results of similar studies.
To visually understand a simulated reionization process, we show the simulated maps of hydrogen number densities, gas temperatures, and ionization degrees at redshifts of 11.1, 7.58, and 5.99 in Fig. \ref{map}. 
This is for the rN256L5RT run, but a similar evolution can be seen in other runs. 
After stars are born at the early epoch of $z\sim 11$, well known "butterfly-shaped" ionised regions associated with star forming regions appear.  
Each ionized region grows, and the half of the simulation volume is ionized at $z\sim 7.5$. 
The reionization is finally completed by $z\sim6$. 

In the top panel of Fig.\ref{ionsfdhis}, we show the simulated ionisation histories. 
As we can see from the figure, in both cases of the high and low resolution run, the mass weighted average ionisation fractions of hydrogen are slightly higher than volume weighted ones. This trend is indicating that high density regions are reionised prior to lower density regions, as in previous studies 
\citep{Iliev07,TC07,Finlator09,Baek09}. 
Obviously, ionised fractions at high redshifts are higher in the high resolution run, since the first UV source tends to appear at higher redshift as the mass resolution gets higher. 

In the bottom panel of Fig. \ref{ionsfdhis}, we represent the simulated star formation rate densities (SFRDs). 
We also plot the SFRDs inferred from observed luminosity functions of high-$z$ drop-out galaxies in the figure \citep{Ouchi09,Bouwens11}. 
It has often been argued that stars are overproduced in numerical simulations. 
Actually our simulations also seem to overproduce stars, if we exclude UV feedback. 
It is interesting that the SFRDs in both runs are suppressed by a factor of $\sim5-6$ at $z=6$ if we include UV feedback. 
This strong suppression makes the  simulated SFRDs match the SFRDs inferred by observations at $z\le8$. 
At $z\sim10$, the simulated SFRDs remain inconsistent with the observation. 
However whether simulations and observations are consistent or not might be inconsequential at this stage, because there are some uncertainty in the both of the numerical and observational studies. 
Concerning our study, as mentioned above, the volume size of the simulation is too small to be regarded as a representative of the Universe. 
Moreover some processes expected to be important for the star formation process, that will be discussed in a later section, haven't been included yet. 
As for observational studies, on the other hand, the shape and amplitude of the luminosity function of faint galaxies are still uncertain especially at high-$z$. 
In addition, the fact that our small simulation box does not contain luminous galaxies that are directly observed illustrates the mismatch between our simulations and the observations.
Thus here the point to which we should pay attention is that UV feedback significantly affects the cosmic star formation history. 
Comparing the results of similar previous studies, such as \cite{Finlator11} and \cite{PS11}, we find that the suppression seen in our simulation is considerably stronger than those shown in these studies.  
We should mention that the former have not appropriately considered the internal UV feedback, while the latter do not have a sufficient mass resolution to discuss the suppression. 
Also none of these two studies take into account $\rm H_2$ chemistry. We study the reasons for our strong suppression of star formation in the following sections. 

In the lower panel of Fig. \ref{ionsfdhis}, we can see that the SFRD in rN256L5RT at $z\le9$ is lower than that in rN128L5RT, in spite of the fact that SFRDs in the corresponding runs without UV feedback, namely rN256L5 and rN128L5, are almost identical at this epoch. 
Relating to the discrepancy of the SFRDs, the ionization history at the later stage in rN256L5RT is delayed compared to that in rN128L5RT, as shown in the upper panel of Fig. \ref{ionsfdhis}.  
These discrepancies are thought to be caused by numerical resolution effects. In Section \ref{ResEf}, we discuss it in detail. 

\subsection{Properties of Halos}
\begin{figure}
	\centering
	{\includegraphics[width=7cm]{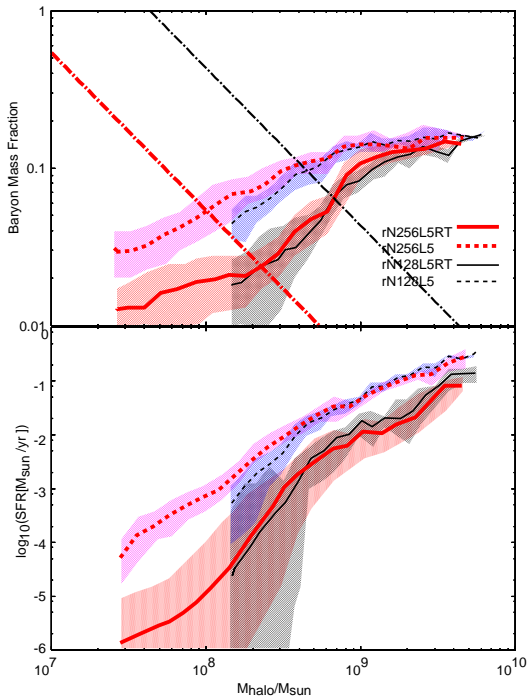}}	
	\caption{Baryon mass fractions (upper panel) and star formation rates (lower panel) at $z=6.0$ as a function of halo mass. 
	The curves are mean values at a given mass range, and the shaded regions attached to the curves are their dispersions. 
	The thick solid, thick dashed, thin solid and thin dashed curves respectively show the results by rN256L5RT, rN256L5,  rN128L5RT and 
	rN128L5. The thick and thin dot-dashed lines in the upper panel indicate the critical baryon mass fraction given by Eq. \ref{fbcr}. 
	Halos located above this lines contain more than 100 baryon particles.}
	\label{fb_halo}
\end{figure}
\begin{figure*}
	\centering
	{\includegraphics[width=14cm]{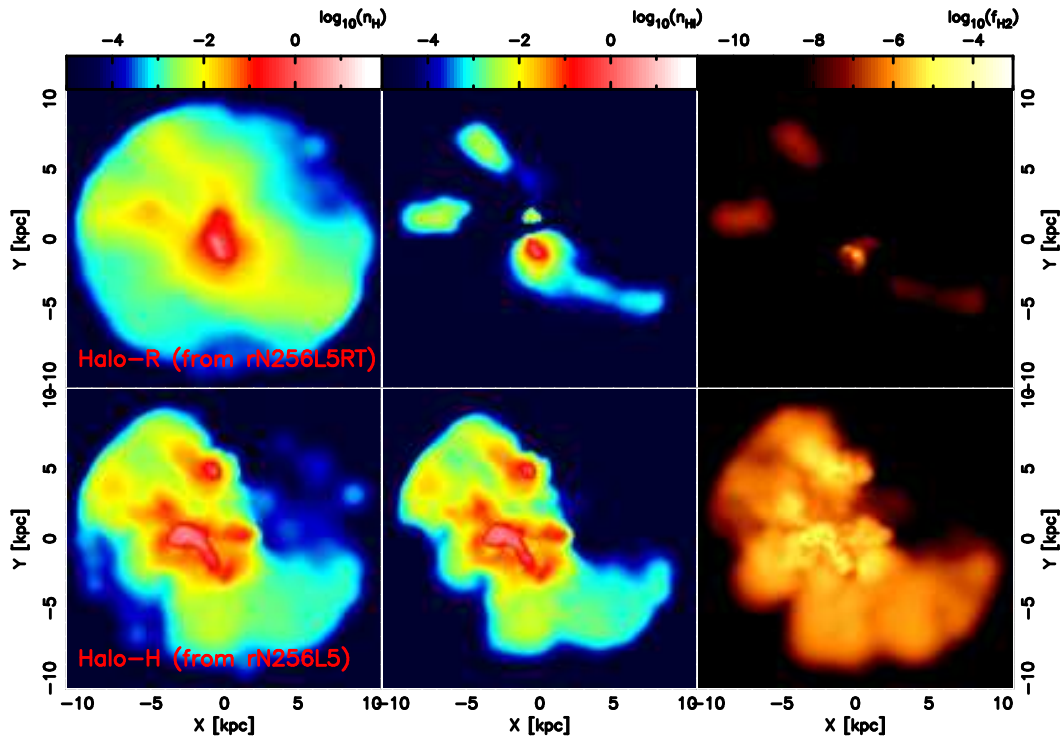}}	
	\caption{Maps of the distributions of $\rm H$ nucleus number density(right panels), neutral $\rm H$ number density(central panels), 
	and $\rm H_2$ fraction(left panels). 
	The upper row is for Halo-R (a massive halo in rN256L5RT run), while the lower row is for Halo-H
	(the same halo but without UV feedback). }
	\label{internal}
\end{figure*}
In the previous subsection, we showed that the cosmic star formation rate is strongly suppressed by UV feedback effects. 
In this section we analyze some properties of the halos to uncover the reasons for the suppressed star formation. 
To identify individual halos, we first use the friends-of-friends method with a linking length of $0.2\times N^{1/3}$. 
Then, gradually shrinking a sphere of given radius, we refine the location of the centre of a halo to be the mass centre of it. When the density within the sphere is equal to the virial density, we define the radius as the virial radius of the halo. 
We define the ratio of the amount of baryon particles within a halo's virial radius to the total halo mass as the baryon mass fraction of the halo. 

Since the suppression of the star formation is the most noticeable at $z\sim6.0$, we here focus on halos at this epoch. 
In the upper panel of Fig.\ref{fb_halo}, we compare the baryon fractions in the rN256L5RT, rN256L5, rN128L5RT, and rN128L5 runs. 
Before we show the results in detail, we should mention the spatial resolution of the presented simulations.
As described above, before the feedback processes start to work, there is a sufficient number of SPH particles to resolve the internal feedback in each halo. However, once the internal feedback starts to work, SPH particles are evacuated from halos. As a result, the spatial resolutions of the hydrodynamics and RT calculations, especially for low mass halos, gradually decreases as time goes by. 
As shown by previous studies, at least 100 SPH particles are required to resolve each halo \citep{PS11,Finlator11}.
The baryon mass fraction which satisfies this critical condition can be given by 
\begin{equation}\label{fbcr}
	f_{\rm b,cr}=5.4\times10^{-2} \left(\frac{10^8M_{\odot}}{M_{\rm halo}}\right), 
\end{equation}
for the high resolution runs. The corresponding equation for the low resolution runs is easily obtained by multiplying the above equation by eight. 
We trace the corresponding lines in the upper panel of Fig.\ref{fb_halo}. Then we can see that  the internal structures of halos with $M_{\rm halo}\le 2\times 10^8M_{\odot}$ (or $M_{\rm halo}\le 7\times 10^8 M{\odot}$ in the low resolution run) cannot be discussed with our simulations. 
Indeed the baryon mass fractions of halos with $M_{\rm halo}=2-5\times 10^8$ in rN128L5RT show larger deviations than those in rN256L5RT does. It is thought to result from a very small number of gas particles in the halos. Therefore we should consider the internal quantities of halos less massive than these mass scales to be unreliable. 

Roughly speaking, the baryon mass fractions of halos with $M_{\rm halo}>10^9M_{\odot}$ are insensitive to UV feedback, though 
the baryon mass fractions are slightly smaller than the value expected from the cosmological parameters $\Omega_{\rm b}/\Omega_{\rm 0}\simeq 0.18$. 
On the other hand, the discrepancy of baryon mass fractions between the runs with and without UV feedback is remarkable for  halos with $M_{\rm halo}< 10^9M_{\odot}$.  
At $z=6.0$, the characteristic mass scale below which photo-evaporation effectively takes place is roughly consistent with the value shown by \cite{Okamoto08}, although they assume an homogeneous external UV feedback and do not consider internal UV feedback and self-shielding. 
This consistency likely comes from the following facts: (i) the temperature of ionised gas is not changed dramatically even if internal UV feedback is included, (ii) for less massive halos, self-shielding is not as efficient as for massive halos \citep{TU98}. 
Therefore, as far as we consider the epoch when the Universe is almost ionised, the characteristic mass investigated with a uniform UV background intensity would be approximately correct. 
But we should mention here that the internal structure  of halos is significantly affected by the internal UV feedback even in massive halos as we will show in Section \ref{Impact}. 

We represent the star formation rate as function of halo mass in the lower panel of Fig.\ref{fb_halo}. 
As expected from the top panel of Fig.\ref{fb_halo}, the star formation rates of halos with $M_{\rm halo} < 10^9M_{\odot}$ is 
strongly suppressed via photo-evaporation. 
An interesting result can be seen in the star formation rates of massive halos. 
The star formation in such massive halos is also suppressed, despite the fact that the massive halos hardly lose any mass. 
In the following two subsections, we show reasons for this phenomenon. 

\subsubsection{Impact of UV feedback on Internal Density Structures} \label{Impact}
A possible mechanism which causes the suppression is the decrease of gas density driven by the internal feedback. 
Soon after the birth of stars in a dense cooled region, the stars emit ionising photons which quickly heat up the surrounding gas up to $\sim 10^4$K. 
This process leads to the destruction of the dense star-forming region by the high thermal pressure. 
Also the propagation of the D-type ionisation front makes blow-out by SN explosions effective \citep{KY05}.
According to the equation (1), the star formation rate is proportional to $\rho^{1.5}$. 
Therefore if the gas density is decreased, this directly indicates that the further star formation in the region is suppressed. 
%Sometimes it is expected that further star formation in a expanding shell occurs. Unfortunately our simulations cannot resolve whether the further star formation in the shell is possible or not 

\begin{figure}
	\centering
	{\includegraphics[width=7cm]{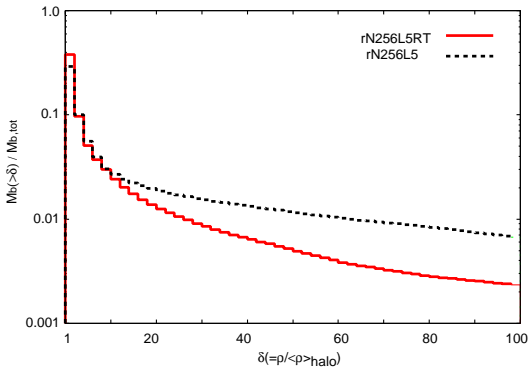}}	
	\caption{Fractions of baryon mass that resides in a place where the baryon overdensity $\delta$ exceeds a given value.
	The solid and dashed curves represent Halo-R and Halo-H respectively. }
	\label{clump}
\end{figure}

We choose one massive halo with $M_{\rm tot} \approx 6.7 \times 10^9M_{\odot}$ from rN256L5RT. 
We also pick the same halo from rN256L5, and compare them in Fig. \ref{internal}. 
Hereafter we call the former Halo-R, and the latter Halo-H. 
In each halo, more than 20,000 SPH particles are included. Thus we can discuss their internal density structures. 
As we can see from Fig. \ref{internal}, the internal density structures of these two halos are quite different. 
Some clumpy structures can be seen in Halo-H, that are significantly smoothened by the UV feedback in 
Halo-R. 
To emphasize the difference in the internal density structures more quantitatively, we show cumulative histograms of the mass fraction where the overdensity $\delta$ exceeds a given value in Fig.\ref{clump}. 
Here we define the overdensity as $\delta \equiv \rho_{\rm b}/<\rho_{\rm b}>_{\rm halo}$, where $<\rho_{\rm b}>_{\rm halo}$ is the average 
gas mass density of the massive halo. 
As shown in Fig. \ref{clump}, Halo-H has more high density components than Halo-R. 
This is the direct cause for the suppressed star formation in the massive halo. 
The destruction of clumpy structures likely works more effectively in less massive halos. 
Since a massive  halo forms through coalescences of smaller halos which are sensitive to the UV feedback in a CDM cosmology, resolving previous episodes of star formation would leads to stronger suppression of the star formation. 
Indeed, in our simulations, the sensitivities of the star formation in massive halos to the UV feedback are higher than those shown by \cite{WC09}. 

It is also expected that the internal UV feedback affects the ionizing photon escape fraction from a galaxy, since the escape fractions strongly depend on the internal structure of a galaxy, such as clumsiness and the covering factor of clumps \citep{FS11}. 
In addition, stars are usually born in a high density region where a high recombination rate is expected. 
Hence if we don't take into account the back-reaction by UV radiation emitted by newly born stars, the ionising photons hardly escape from the star forming regions.  
In \cite{Ume12}, we have explored the difference between the ionizing photon escape fractions from Halo-R and from Halo-H, by solving the RT of ionising photons in post-processing. 
As a result, we have found that the escape fractions of the HI, HeI, and HeII ionising photons from Halo-R are respectively 0.31, 0.35, and 0.13. 
These values were indeed higher that those from Halo-H by a factor of 2. 

%According to the limited mass resolution, our simulations cannot resolve low mass gas clouds where stars are born in the realistic case. 
Our mass resolution does not allow as to resolve the structure of the low mass clouds where stars are born. 
Therefore we should consider the possibility that the results presented in this section would be altered if we increase the numerical resolution. 
Recent high resolution RHD simulations by \cite{Dale12} have shown that the dynamical effect of photo-ionisation disrupts low-mass star-forming clouds with $10^{4-5}M_{\odot}$ within a few Myr. 
This result indicates that further star formation in these regions is suppressed and the feedback enhances the escape fractions of ionizing photons from the star-forming clouds. 
Thus we expected that the results in this section are qualitatively unaffected by the numerical resolution, even though the quantitative values might be affected.  

\subsubsection{Dissociation of $\rm H_2$ molecules}\label{h2}
In previous studies of reionisation by numerical simulations, the non-equilibrium chemistry of $\rm H_2$ has often been neglected for simplicity. 
As mentioned above, if we want to resolve the internal structure of halos, which is closely related with the star formation rate and with the escape probability of ionising photons, we have to consider cooling processes below $10^4$K. 
In the low metallicity environment which is prevalent in the early Universe, $\rm H_2$ molecules are the most efficient coolants in the low temperature regime. 
Therefore, if the $\rm H_2$ fraction is considerably decreased, the gas cannot cool down to $T <10^4$K and stars are hardly formed. 

\begin{figure}
	\centering
	{\includegraphics[width=7cm]{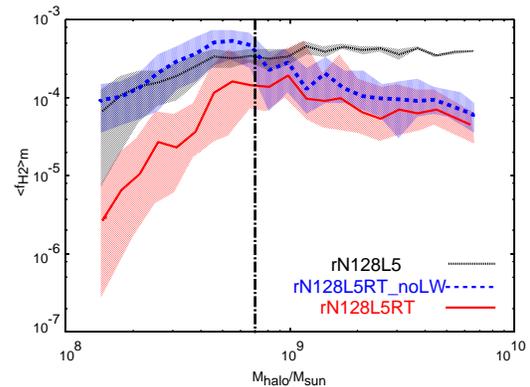}}	
	\caption{Mass weighted average $\rm H_2$ fractions in high density regions ($\delta \ge 1$) of halos at $z=6$ 
	as a function of halo mass. 
	The dotted, dashed and sold lines show the results of rN128L5, rN128L5RT-noLW and rN128L5RT, respectively.
	The vertical dot-dashed line indicates the mass of halo which typically includes 100 SPH particles in the rN128L5RT run.}
	\label{fH2}
\end{figure}
\begin{figure}
	\centering
	{\includegraphics[width=7cm]{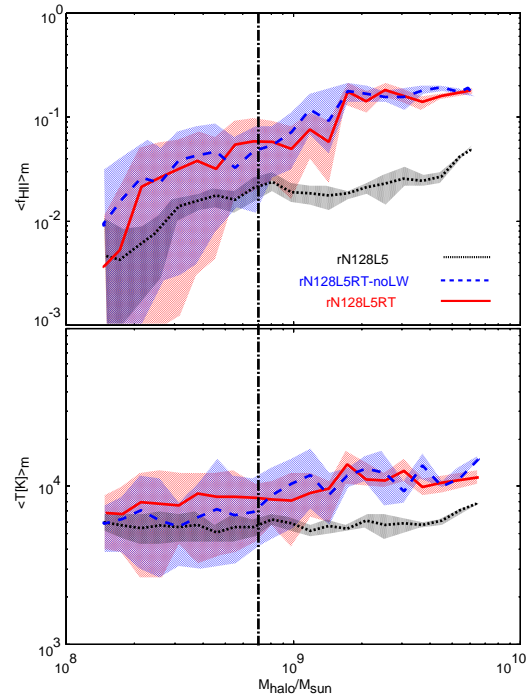}}	
	\caption{Ionised hydrogen fractions (upper panel) and temperatures(lower panel) in high density regions ($\delta \ge 1$) 
	of halos at $z=6$ as a function of halo mass. 
	The dotted, dashed and sold lines show the results of rN128L5, rN128L5RT-noLW and rN128L5RT, respectively.
	The vertical dot-dashed line indicates the mass of halo which typically includes 100 SPH particles in the rN128L5RT run.}
	\label{TfHII}
\end{figure}
In this section, we show how severely the $\rm H_2$ fraction of each halo is affected by UV feedback, i.e., the photo-ionisation and photo-dissociation processes. 
%The interpretation of the results shown in this section is not straightforward. 
Before we show the results in detail,  we briefly introduce some processes closely related to the determination of $\rm H_2$  fraction before we discuss the results. 
The destruction of $\rm H_2$ molecules is  caused not only by photo-dissociating radiation but also indirectly by photo-ionising radiation. 
In the photo-ionised hot regions, $\rm H_2$ molecules are collisionally dissociated. Moreover, the smoothing of internal density structures shown in the previous section weakens the self-shielding effect. 
On the other hand, there is a process which promotes $\rm H_2$ formation. In the absence of dust grains, $\rm H_2$ molecules are produced via the $\rm H^-$ process in which electrons work as catalysts. Thus as the $\rm e^-$ fraction increases, $\rm H_2$ formation rate also increases with a given temperature. Actually, the promotion of $\rm H_2$ formation always works when an enhancement of the electron fraction exists, but usually the positive feedback is remarkable only if the photo-ionisation rate is not too high and the medium is only slightly ionized. Otherwise the strong negative photo-ionisation feedback becomes dominant. 
%We show which of the processes dominantly works in each halo mass range in the following. 

As already mentioned, the internal quantities of low mass halos are not reliable due to the very small number of gas particles in the halos. 
Therefore we will only discuss the internal quantities in massive halos with masses $\ge 10^9M_{\odot}$ about which we are currently curious.
In addition, we here focus only on high density regions, since stars are expected to form there. 
We show the mass weighted average $\rm H_2$ fraction in the high density regions of halos at $z=6.0$ in Fig.\ref{fH2}. 
To emphasize the effects of photo-ionisation, we also plot the mass averaged ionised fraction and temperature in high density regions in Fig. \ref{TfHII}. 
We define high density regions as the places where $\delta \ge 1.0$, where the definition of $\delta$ is the same as in the previous section. 
We show not only the $\rm H_2$ fraction in the rN128L5 and rN128L5RT runs but also in the rN128L5RT-noLW run, in which the photo-dissociation and photo-detachment processes are excluded, in order to evaluate impact of the photo-dissociating radiations. 

In the massive halos, the $\rm H_2$ fractions in both the rN128L5RT and rN128L5RT-noLW runs are considerably smaller than in the rN128L5 run.  Moreover, the difference in the $\rm H_2$ fractions between in the rN128L5RT and rN128L5RT-noLW runs is not so large. 
This result tells us two facts. First, star formation in massive halos is inhibited by the destruction of $\rm H_2$ molecules. 
Second, the destruction of $\rm H_2$ is caused by photo-ionisation(-heating) more than by photo-dissociation. 
As obviously shown in the upper panel Fig. \ref{TfHII}, the mass weighted ionised fraction tends to increase as the halo mass increases, though the ionisation degrees are rather small. 
This trend basically originates in the fact that the specific star formation rate becomes higher as the halo mass increases, as shown by the bottom panel of Fig \ref{fb_halo}. 
A similar trend can also be seen in the average temperatures of halos. 
The gas in massive halos is heated up to $\sim 8,000-10,000$K, in spite of the fact that they are not so highly ionised. 
It has been known that such a mild photo-heating is caused by high energy photons. 
High energy photons can travel longer paths than lower energy photons, since the cross-sections for photoionisation processes become smaller as the energy of incident photons increases (equations 3, 4, and 5). In addition, the heating rate by one high energy photon is higher than that by a lower energy photon, since the released electron has a larger kinetic energy. As a result, the gas is mildly heated up even if it is almost neutral. 
Since star formation in massive halos is continuous, the average $\rm H_2$ fractions in these halos would take an almost-equilibrium value.
Indeed, the average $\rm H_2$ fractions in the massive halos, about $10^{-5}-10^{-4}$ as shown in Fig. \ref{fH2}, are roughly consistent with the equilibrium value at $\sim 8,000-10,000$K with $f_{\rm HII}\approx0.1-0.2$.  
Here we should mention that such mildly heated regions are thought to be the places where the star formation is originally possible in rN128L5 run. 
It is also true that photo-dissociating radiation can affect higher density regions than photo-ionising radiation can, but the amount of gas residing in the regions solely affected by the photo-dissociating radiation are smaller than that in the mildly heated up regions. 
Hence we conclude that the suppression of star formation in massive halos is mainly caused by the mild photo-heating and the resultant collisional dissociation in addition to the destruction of clumpy structures shown in the previous section.   

If we focus on halos with masses $\sim 5\times 10^8M_{\odot}$, we notice an interesting fact that the $\rm H_2$ fractions in the rN128L5RT-noLW run are slightly higher than those in the rN128L5 run. This trend might result from the positive feedback due to the enhancement of the electron fraction, but we consider these halos not to be sufficiently resolved to draw this conclusion with any certainty. 

\section{Discussion}
\subsection{Dependence of the Results on Treatment of Photo-dissociation}
\begin{figure}
	\centering
	{\includegraphics[width=7cm]{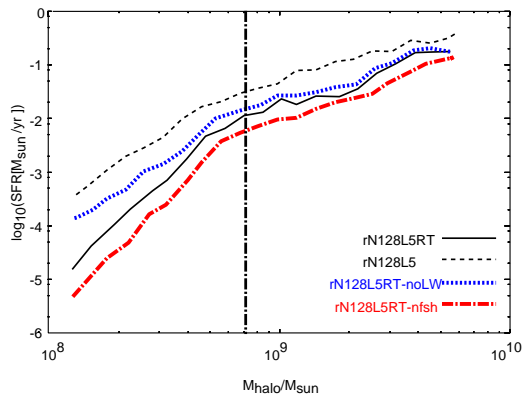}}	
	\caption{Star formation rates of halos at $z=6$ shown as a function of halo mass. 
	The solid, dashed, dotted, and dot-dashed curves indicate the results 
	in the rN128L5RT, rN128L5, rN128L5-noLW, and rN128L5-nfsh runs, respectively. 
	The vertical dot-dashed line indicates the halo mass which typically includes 100 SPH particles in the rN128L5RT run.}
	\label{fs_halo}
\end{figure}
\begin{figure}
	\centering
	{\includegraphics[width=7cm]{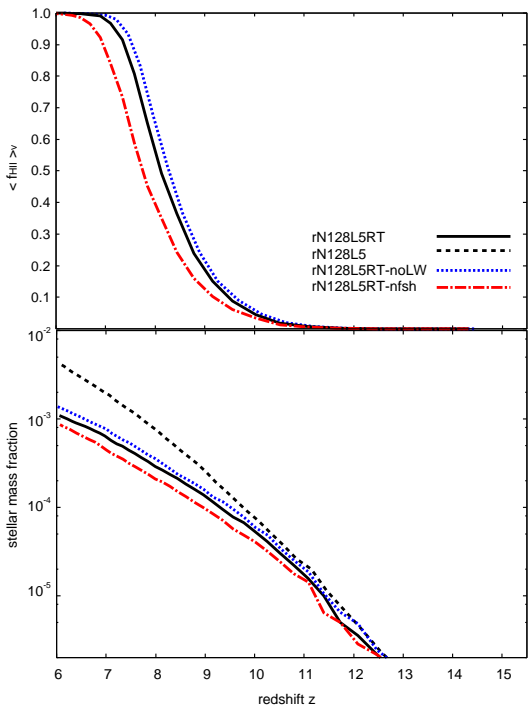}}	
	\caption{Evolution of volume weighted average HII fractions (top panel) and stellar mass fractions (lower panel) 
	shown as a function of redshift. 
	The solid, dashed, dotted, and dot-dashed curves indicate the results 
	in the rN128L5RT, rN128L5, rN128L5-noLW, and rN128L5-nfsh runs, respectively. }
	\label{fs_kd}
\end{figure}
As we already mentioned in Section \ref{feedback}, the self-shielding function we employ might lead to a stronger shielding effect than if  we correctly solve the transfer of LW band photons. 
Hence we should analyse how our results change with the treatment of the photo-dissociating process. 
In Fig.\ref{fs_halo}, we show the star formation rate of halos. 
As simply expected from Fig. \ref{fH2}, in massive halos, the star formation rates in rN128L5RT and rN128L5RT-noLW are nearly equivalent. 
On the other hand, for less massive halos, the star formation rates in rN128L5RT-noLW are slightly higher than that in rN128L5RT due to the larger fraction of $\rm H_2$ molecules. 
We also plot the star formation rates of the rN128L5RT-nfsh run. 
Since we make the optically thin approximation for LW radiations in the rN128L5RT-nfsh run, the photo-dissociation process is maximal in this run. 
We find that the discrepancy between the star formation rates in rN128L5RT and rN128L5RT-nfsh runs is a factor of $\sim 1.5-2.0$ at most.  
We also show the resultant reionisation histories in Fig.\ref{fs_kd}. 
We find that the difference of the overlapping redshifts, when the average neutral fraction is 0.5, in the rN128L5RT and rN128L5RT-nfsh runs is just $\Delta z\approx 0.4$ corresponding to $\Delta \tau_{\rm e}\approx 0.036$. 
Therefore we expect that our results would not change dramatically if we solved correctly the transfer of LW photons. 

\subsection{HI Fraction at the End of Reionzation and Its Dependence on the Choice of ISM Escape Fraction}\label{escape}
\begin{figure}
	\centering
	{\includegraphics[width=7cm]{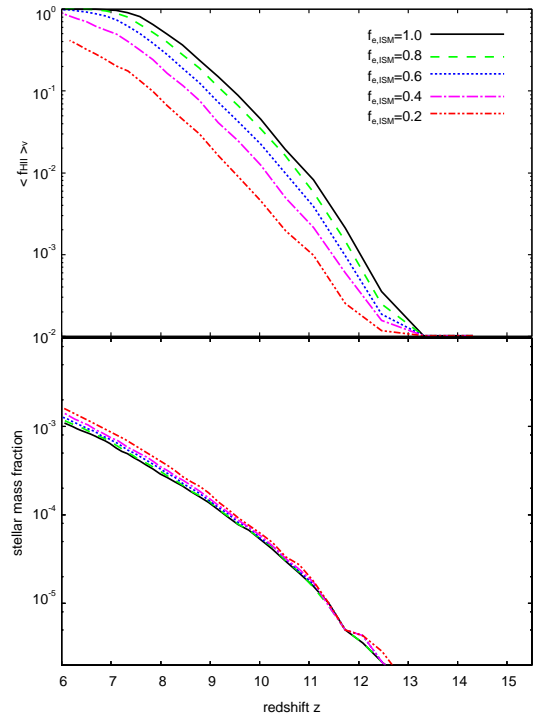}}	
	\caption{Evolution of the volume weighted average HII fraction (upper) and stellar mass fraction (lower). 
	The results computed with $f_{\rm e,ISM}=1.0$ (solid curve), $f_{\rm e,ISM}=0.8$ 
	(dashed curve), $f_{\rm e,ISM}=0.6$ (dotted curve), $f_{\rm e,ISM}=0.4$ (dot-dashed curve), 
	and $f_{\rm e,ISM}=0.2$ (dot-dot-dashed curve) are shown.}
	\label{fs_fe}
\end{figure}
\begin{figure}
	\centering
	{\includegraphics[width=7cm]{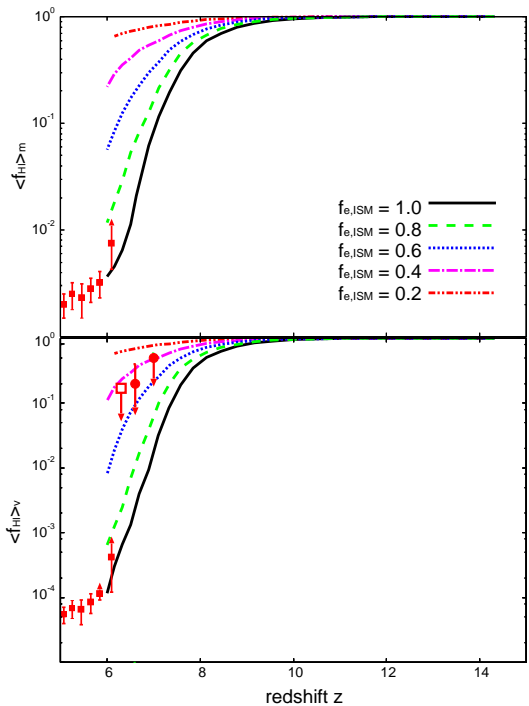}}	
	\caption{The evolution of the volume and mass weighted average HI fractions is shown in the upper and bottom panels, respectively. 
	The results computed with $f_{\rm e,ISM}=1.0$ (solid curve), with $f_{\rm e,ISM}=0.8$ 
	(dashed curve), with $f_{\rm e,ISM}=0.6$ (dotted curve), with $f_{\rm e,ISM}=0.4$ (dot-dashed curve), 
	and with $f_{\rm e,ISM}=0.2$ (dot-dot-dashed curve) are plotted.
	The open square is the upper limit estimated from the constraints from $\rm Ly_{\alpha}$ damping wing of GRB at $z=6.3$ (Totani 
	et al. 2006). 
	The filled circles are the upper limits estimated from observations of LAEs at $z=6.6$ (Ouchi et al. 2010).
	The filled squares are observational values evaluated with the spectra of QSOs at $z\sim6$ (Fan et al. 2006).}
	\label{fHI}
\end{figure}
In this study, we never assumed a "galactic" escape fraction, since our calculation automatically takes the absorption within a halo into account. 
We also assumed no photon losses in the subgrid, unresolved medium, and so worked with an ISM escape fraction of $f_{\rm e,ISM}=1.0$. 
But unfortunately there is no guarantee that our simulations sufficiently resolve absorption in the ISM. 
Hence we first show how our results change with different choices of $f_{\rm e,ISM}$. 
We show ionisation histories simulated with $f_{\rm e,ISM}=0.2-1.0$ in the upper panel of Fig.\ref{fs_fe}. 
As shown in the figure, ionisation history strongly depends on the choice of $f_{\rm e,ISM}$. 
Obviously, reionisation proceeds more quickly as $f_{\rm e,ISM}$ increases. 
Here we should point out an interesting fact;  
in spite of the strong dependence of the reionisation history on $f_{\rm e,ISM}$, the star formation history is almost independent on $f_{\rm e,ISM}$ as we can see in the bottom panel of Fig. \ref{fs_fe}. 
This implies that the global star formation at such high redshifts is mainly regulated by the internal feedback processes rather than the  external UVB, at least in our simulations. 
Note however that it might not be always true, since our simulation volume is too small to be representative of the Universe. As shown in \cite{Iliev06b}, the global reionisation and star formation histories vary considerably such small volumes. Moreover we cannot consider UV radiation emitted from distant sources properly because of the limited volume. 

In Fig.\ref{fHI}, we plot the simulated neutral hydrogen fractions and compare them with neutral hydrogen fractions observationally estimated from the spectra of 19 QSOs at $z\sim6$ \citep{Fan06b}, from Ly$\alpha$ damping wing of Gamma-ray burst 050904 at $z=6.3$ \citep{Totani06}, and from the evolution of the LAE luminosity function at $z=6.6$ \citep{Ouchi10}. 
As shown in the figure, our simulation results and observed neutral hydrogen fractions are well consistent if we assume $f_{\rm e,ISM}$ greater than $\sim 0.5$, otherwise simulated neutral fractions are larger than the observational values. 
The ISM escape fraction should approach unity as the mass resolution increases.  
Unfortunately it is still unclear how much resolution we need to remove such an uncertainty. 
But we think $f_{\rm e,ISM}>0.5$ for our simulations is reasonable, because of the following arguments. 
\cite{WC09} have shown that an ionizing photon escape fraction from a galaxy with mass of $\sim 10^8M_{\odot}$, which roughly corresponds to the minimum halo mass in our low resolution run (rN128L5 series), is $0.4-0.7$.  The maximum spatial resolution in their simulations is 0.1pc so that the clumpiness in a galaxy is possibly sufficiently resolved. 
\cite{WC09} have also commented that the computed escape fractions should be considered as lower limits, since their simulations have not taken into account previous episodes of star formation which likely decreases the baryon mass fraction as shown in this paper. 
In addition, needless to say that the galactic escape fraction should be smaller than the ISM escape fraction since the former includes the latter.
Moreover, \cite{Dale12} have shown the escape fractions from clouds with $10^{4-5}M_{\odot}$, which roughly corresponds to the mass resolution of our simulations, are typically greater than 0.6. 

In \cite{Ume12}, we have computed the escape fractions of HI, HeI, and HeII ionising photons from the massive halo shown in section \ref{Impact}. The computed escape fractions are 0.31, 0.35, and 0.13 for HeI, HeI, and HeII ionising photons, respectively. 
However we still have many unknowns, e.g., the sufficient mass resolution to evaluate the escape fraction, a typical value of the escape fraction of a halo with a given mass, the dependence of the escape fraction on halo mass and/or redshift, and so on. 
We will do a statistical study on the escape fraction to clarify such unknowns in the future. 

\subsection{Optical Depth for Thomson Scattering}
\begin{figure}
	\centering
	{\includegraphics[width=7cm]{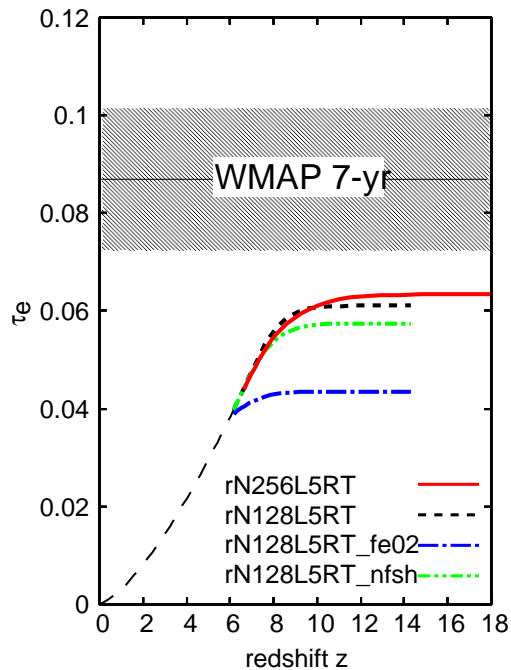}}	
	\caption{Optical depths for the Thomson scattering. The solid, dashed, dot-dashed, and dot-dot-dashed, 
	curves respectively show the optical depths evaluated from the rN256L5RT, rN128L5RT, 
	rN128L5-fe02, and rN128L5RT-nfsh runs. }
	\label{tau}
\end{figure}
The measurement of the optical depth for the Thomson scattering of CMB photons by electrons released during the reionisation is frequently applied  for evaluating the reionisation epoch. 
The WMAP 7year have shown that the optical depth is $\tau_{\rm e}=0.087\pm0.014$ , which corresponds a reionisation at $z_{\rm r}=10.4\pm1.2$ under the assumption of instantaneous reionisation. 
In this section, we present the optical depth computed from the simulated reionisation history. 
It is likely invalid to compare the optical depth computed from our results with the observational value, since the volume size of the simulations is insufficient.  
However comparing the optical depths computed from  different runs is useful to understand the simulation results. 
%Moreover someone who carries out similar simulations would compare the results. 
We evaluated the optical depths in our simulations by 
\begin{equation}
	\tau_{\rm e} = c\sigma_{\rm t}\int ^0_{z_r} n_{\rm e}(z)\frac{dt}{dz}dz, 
\end{equation}
where $c$, $\sigma_{\rm t}$, and $n_{\rm e}(z)$ are the speed of light, the corresponding cross-section, and the electron number density at a given redshift, respectively. 
Since our simulations are stopped at $z=6.0$, we cannot evaluate electron fractions at $z<6.0$ with the simulations. 
Therefore, to compute the optical depth, we assumed that the H and He in the IGM at $z<6.0$ is fully ionised. 
Note that we probably overestimate the optical depth at $z<6.0$ with the assumption, since HeII reionization was expected to take place at $z\sim3$.
In Fig. \ref{tau}, we show the optical depths computed in several runs , and compare them with the WMAP result. 
It is obvious that the optical depths of all runs never reach the observed value. 
Since the first UV sources would appear at higher redshift as the mass resolution increases, we simply speculate that the total optical depth is higher in a higher resolution run. 
As we can see in Fig. \ref{tau}, the above speculation is indeed correct, but the increment of the optical depth is small.  
This originates in the fact that the ionisation history in rN256L5RT run is delayed by UV feedback during the later stage (see Fig. \ref{ionsfdhis}). 

As we can see immediately from Fig. \ref{tau}, any the optical depths estimated from our results are not consistent with observed value. 
However, no conclusions can be drawn from this, since our simulation volume is not sufficient to directly compare the optical depth. 
In addition our simulations didn't include Pop III stars which are expected to have a large contribution to the early stage of the reionsation\citep{Ciardi03,Sokasian04,Ahn12}.

\subsection{Chemical Feedback}
In this paper, chemical feedback by SN explosions was not included. The metal enrichment likely plays an important role. The increase of metal directly leads to an enhancement of the cooling rate owing to additional cooling processes. 
Moreover, $\rm H_2$ molecules are efficiently produced on the surface of dust grains. 
This also leads to an enhancement of cooling rate below $10^4$K.  
Not only these effects on cooling rates, but also the absorption of UV photons by grains are closely related to the cosmic star formation and reionisation history. 
For instance a direct consequence is that the number of ionising photons escaping from halos decreases. 
The absorptions of UV photons by grains would alleviates UV feedback, but push the matter outward by the radiation force. 
Hence the net effect of metal enrichment on the star formation and reionisation history is still uncertain. 
Moreover, it is expected that the metal enrichment process regulates the transition from Pop III to Pop II star formation mode. 
We will study the complicated role of the metal enrichment in a forthcoming paper (Paper II).  

\subsection{Resolution Effects}\label{ResEf}
In this section, we discuss how numerical resolution affects our results.  
As we can see from the lower panel of Fig. \ref{ionsfdhis}, our simulation result have not converged; The cosmic star formation rates at the later stage in the high resolution rN256L5RT run is lower than those in the low resolution rN128L5RT run. 
\cite{SU04} have shown that a considerable fraction of the gas in a small galaxy is evaporated by UVB, but the star formation in high density regions is still possible owing to the self-shielding effect. 
Since our simulations cannot resolve some small clouds where the self-shielding effect originally works, galaxies in the low resolution run are thought to be more sensitive to the UV feedback than those in the high resolution run. 
Due to this type of resolution effect, we might underestimate the baryon mass fractions and the star formation rates especially in low mass galaxies. However this resolution effect cannot explain why the star formation rates at $z\le9$ in the high resolution run is lower than those in the low resolution run. 

There is another type of resolution effects in our simulations. 
As we increase the mass resolution, the epoch when the first source forms is shifted towards higher redshifts, since our numerical resolution is not enough to resolve Pop III halos which are the first collapsed objects in the Universe. 
This fact indicates that the feedback processes start to work from an earlier epoch in the higher resolution run. 
Since a massive galaxy forms via coalescences of less massive galaxies which are already damaged by previous feedback episodes, 
the star formation in the high resolution run is likely delayed compared to that in the low resolution run. 
This trend caused by the latter type of resolution effect is basically consistent with the trend shown in Fig.\ref{ionsfdhis}. 
Therefore it is very important to increase the numerical resolution up to resolving the the scale of Pop III halo $M \sim 10^{5-6}M_{\odot}$ \citep{Yoshida03}, not only for understanding the contribution of Pop III  halos to the evolution of the Universe, but also for removing this type of resolution effect. 

\section{Conclusions}
We explored the impact of UV feedback on reionisation history using RHD simulations in which star formation regulated by the internal and external UV feedback in each galaxy is appropriately considered. 
As a result we found that the cosmic star formation rate is considerably decreased due to the UV feedback.  
In  halos with $M<10^9M_{\odot}$, the star formation is mainly suppressed by photo-evaporation and photo-dissociation. 
Interestingly, more massive halos also suffers from a strong suppression of star formation, although they hardly lose their baryon components by photo-evaporation. 
For massive halos, the suppression by photo-dissociation is not very effective, but the high density gas clumps are significantly damaged by the internal UV and SN feedback. 
In addition, mild photo-ionisation in the massive halo turns out to keep the gas hot, and consequently $\rm H_2$ molecules are destroyed via collisional dissociation. 
We also found that the degree of the suppression is larger in higher resolution run, because the UV feedback works since an earlier stage, and small galaxies that can be resolved only in the high resolution run are more sensitive to the UV feedback. 
Therefore the reionisation process turns out to be delayed at $z\le9$ in the high resolution run, compared to that in the low resolution run. 
This trend is indeed a kind of the resolution effect, but this fact is indicating that resolving Pop III halos is essential to compute the correct star formation and reionisation histories. 

We also find that the simulated cosmic star formation rate densities and HI fractions at $z\le7-8$ are well consistent with those inferred from observations, even though our simulation volume is too small to be a representative patch of the Universe. 
%However the optical depths for Thomson scattering of CMB photons computed from our simulations never reach the corresponding value observed by WMAP. These results imply that some additional ionising sources, such as Pop III stars, dominantly contribute to the cosmic reionisation process especially at the early Universe.  

\section*{Acknowledgements}
We are grateful to the members of LIDAU meetings held in Paris and Strasbourg for fruitful discussion. 
We also thank to the anonymous referee for valuable comments. 
KH thank M. Umemura, T. Okamoto, and K. Yoshikawa for useful comments on the numerical method. 
Numerical simulations were performed with T2K-Tsukuba at Centre for Computational Sciences in University of Tsukuba, with Cray XT4 at CfCA of NAOJ, and with Blue Gene/P Babel at the CNRS computing center: IDRIS. 
This work was supported in part by Grant-in-Aid for Young Scientists B: 24740114, by MEXT HPCI STRATEGIC PROGRAM, and by the french funding agency ANR (ANR-09-BLAN-0030).

\label{lastpage}

\end{document}